\documentclass[12pt,preprint]{aastex}
\usepackage{epsf,color}

\newcommand{\iphj}{{i+\frac{1}{2},j}}
\newcommand{\ijph}{{i,j+\frac{1}{2}}}
\newcommand{\imhj}{{i-\frac{1}{2},j}}
\newcommand{\ijmh}{{i,j-\frac{1}{2}}}

 \newcommand{\rhozero}{\rho_0}

\newcommand{\sfrac}[2]{\mathchoice
  {\kern0em\raise.5ex\hbox{\the\scriptfont0 #1}\kern-.15em/
   \kern-.15em\lower.25ex\hbox{\the\scriptfont0 #2}}
  {\kern0em\raise.5ex\hbox{\the\scriptfont0 #1}\kern-.15em/
   \kern-.15em\lower.25ex\hbox{\the\scriptfont0 #2}}
  {\kern0em\raise.5ex\hbox{\the\scriptscriptfont0 #1}\kern-.2em/
   \kern-.15em\lower.25ex\hbox{\the\scriptscriptfont0 #2}}
  {#1\!/#2}}

\newcommand{\half}{\sfrac{1}{2}}
\newcommand{\myhalf}{\sfrac{1}{2}}
\newcommand{\nph}{{n + \myhalf}}
\newcommand{\nmh}{{n - \myhalf}}

\newcommand{\uadv}{\mathbf{U}^{\mathrm{ADV}}}
\newcommand{\V}{\mathbf{V}}
\newcommand{\Vx}{\mathrm{V}^x}
\newcommand{\Vr}{\mathrm{V}^r}
\newcommand{\uadvstar}{\mathbf{U}^{\mathrm{ADV,*}}}
\newcommand{\wadv}{w^{\mathrm{ADV}}}
\newcommand{\dt}{\Delta t}
\newcommand{\dr}{\Delta r}
\newcommand{\dx}{\Delta \x}

\newcommand{\ubold}{\mathbf{U}}
\newcommand{\ut}{\widetilde{\ubold}}

\newcommand{\er}{\mathbf{e}_r}

\newcommand{\x}{\mathbf{x}}
\newcommand{\rp}{r^\prime}

\newcommand{\gcc}{\mathrm{g~cm^{-3} }}

\newcommand{\ext}{\mathrm{ext}}

\begin{document}

\title{Low Mach Number Modeling of Type Ia Supernovae. II. Energy Evolution}

\shorttitle{Low Mach Number Modeling of SNe Ia}
\shortauthors{Almgren et al.}

\author{A.~S.~Almgren\altaffilmark{1},
        J.~B.~Bell\altaffilmark{1},
        C.~A.~Rendleman\altaffilmark{1},
        M.~Zingale\altaffilmark{2,3}}

\altaffiltext{1}{Center for Computational Science and Engineering,
                 Lawrence Berkeley National Laboratory,
                 Berkeley, CA 94720}

\altaffiltext{2}{Dept. of Astronomy \& Astrophysics,
                 The University of California, Santa Cruz,
                 Santa Cruz, CA 95064}

\altaffiltext{3}{Dept. of Physics \& Astronomy,
                 SUNY Stony Brook,
		 Stony Brook, NY 11794-3800}

\begin{abstract}

The convective period leading up to a Type~Ia supernova (SN~Ia)
explosion is characterized by very low Mach number flows, requiring
hydrodynamical methods well-suited to long-time integration.  We
continue the development of the low Mach number equation set for
stellar scale flows by incorporating the effects of heat release
due to external sources.  
Low Mach number hydrodynamics equations 
with a time-dependent background state are derived, and a numerical method 
based on the approximate projection formalism is presented.  We demonstrate
through validation with a fully compressible hydrodynamics code  
that this low Mach number model accurately captures the expansion of the 
stellar atmosphere as well as the local dynamics due to external heat sources.
This algorithm provides the basis for an efficient simulation tool 
for studying the ignition of SNe~Ia.

\end{abstract}


\section{Introduction}

Modeling the period of convection leading up to the ignition of 
Type~Ia supernovae (SNe~Ia) is critical to determining the distribution of
hot spots that seed the subsequent explosion.  Multidimensional
simulations of SNe~Ia explosions presently seed one or more hot spots
at or near the center of a white dwarf and use a flame model to
describe the subsequent evolution (see for example
\citealt{roepke2005,gamezo2005,plewa:2004}).  Variations in the size,
number, and distribution of these seeds can lead to large differences
in the explosion outcome
\citep{niemeyer:1996,garciasenz:2005,livne:2005}.  The convective
flows leading up to ignition have Mach numbers of 0.01 or less, with
temperature perturbations of only a few percent
\citep{woosley2001,Woosley:2004}---conditions that are extremely
challenging for fully compressible codes.  One-dimensional statistical
methods \citep{wunschwoosley:2004} predict that off-center ignition is
likely, but they cannot give information about the distribution of the
hot spots.  Only recently has progress been made in multidimensional
modeling of convection in white dwarfs.  The first such calculations
\citep{hoflichstein:2002} evolved a two-dimensional wedge of the star
for a few hours using an implicit hydrodynamics algorithm.  Convective
velocities of approximately 100~km~s$^{-1}$ developed.  They observed
compression near the center of the star leading to slightly off-center
ignition.  Three-dimensional anelastic calculations
\citep{kuhlen:2005}, showed a large-scale dipole flow dominating the
evolution, leading to an off-center ignition.  These simulations used
a spectral decomposition, with a small portion of the center of the
star removed due to the coordinate singularity at $r=0$.  Neither of
these calculations operated at a Reynolds number large enough to see
fully developed turbulence.  Further three-dimensional studies are
needed to see how robust this dipole flow is to rotation and
convection at higher Reynolds and Rayleigh numbers.

The goal of the present work is the development of a new
multidimensional hydrodynamics algorithm capable of evolving the full
star from the convective phase, though ignition and into the early
stages of flame propagation.  Long time integration is critical.  As
we showed previously (\citealt{ABRZ:I}---henceforth paper~I), the low
Mach number hydrodynamics equations provide an accurate description of
flows with Mach numbers less than 0.2.  By filtering out sound waves,
the low Mach number approximation allows for much larger time steps
($\sim 1/M$ larger) than corresponding compressible codes.  In
contrast to the anelastic equation set, the low Mach number equations
are capable of modeling flows with finite-amplitude density and
temperature perturbations.  The only restriction is that the pressure
perturbation be small.  Furthermore, because the compressibility
effects due to both the background stratification and local heat
release are included, the low Mach number equations set can
self-consistently evolve the expansion of a hydrostatic atmosphere due
to heat release \citep{almgren:2000}.

In this paper, we continue the development of the low Mach number
hydrodynamics algorithm to include the effects of heat release.  An
energy equation is added, and an earlier assumption from paper~I
is relaxed, now allowing the background state to vary in time.  
In \S~\ref{sec:lmn} we develop the low Mach number equation
set.  In \S~\ref{sec:numer} the numerical methodology is explained.
Comparisons to fully compressible calculations are provided in
\S~\ref{sec:results} to demonstrate the accuracy and utility of our
new algorithm.  We conclude in \S~\ref{sec:conclusions}.

\section{Low Mach Number Hydrodynamics}
\label{sec:lmn} 

In paper~I, we derived a system of low Mach number equations for stellar atmospheres
for which there was a time-independent background state.   The necessary assumption 
for validity of this system was that the Mach number ($M$) of the flow be
small.  In this case, any pressure deviations from the base state pressure,
which are $O(M^2),$ are also small. The perturbations of density and temperature
need not be small.

This system is valid for many low Mach number terrestrial and stellar flows, 
but fails to capture the correct atmospheric response to 
large-scale heating that radially shifts the entire atmosphere at and
above the level at which the heating occurs. This was shown
analytically for the terrestrial atmosphere, using the pseudo-incompressible
approximation, in \cite{bannon:1996a}.  In \citet{almgren:2000} it was
shown that when time variation of the background state is correctly
included, the solution calculated by the low Mach number equation set 
is identical to that reached by the fully compressible equation set.

Physically, this is consistent with the interpretation of the low Mach number
equations as representing instantaneous acoustic equilibration.  If heating
of a local parcel of fluid results in a large temperature perturbation from the ambient, 
the effect of the resultant acoustic waves is to return the parcel to pressure
equilibrium with the fluid around it by expansion of the parcel. The 
density and temperature variations of the parcel relative to the ambient values
may be large, but the parcel will remain close to pressure equilibrium.

By contrast, when an entire layer of the atmosphere is heated, 
the acoustic equilibration process brings the entire layer to a new hydrostatic 
equilibrium.  Consider a horizontally uniform atmosphere with heating uniformly
applied throughout a horizontal layer.   The response of the atmosphere
will itself be horizontally uniform, and for positive heating 
each parcel of fluid above the heated layer will rise in its respective 
radial column.  
In equilibrium, given the assumption that no fluid is lost 
at the top boundary, then following an upward shift of each parcel,
the mass of fluid above any given parcel will not have changed. 
If gravitational acceleration is effectively constant
over the length scale of the base state displacement, then the weight 
of fluid above a parcel will not have changed. Thus for an atmosphere in 
hydrostatic equilibrium the pressure of each parcel will not have changed, 
although that parcel will have changed its radial location.   In other words, 
the material derivative, rather than the time derivative, of the base state pressure 
must be zero.  Numerical examples in \S~\ref{sec:results} of this paper will confirm
the time-dependent response of the base state as well as of the full state is necessary to
correctly capture the atmospheric response to large-scale heat release.

We note that the assumption of constant gravity over the displacement
distance of the base state does not mean one must 
assume constant gravity over the full domain.
In practice the length scale of the base state 
adjustment is much smaller than the length scale of more localized motions.

We now generalize the low Mach number equation set from paper~I
to allow for time dependence of the base state. 
We recall from paper~I the fully compressible
equations of motion in a stellar environment, but here we add an
external heat source, $H_{\ext}$, and again neglect compositional and
reaction terms:
\begin{eqnarray}
        \frac{\partial \rho}{\partial t} + \nabla \cdot (\rho \ubold)
        &=& 0 \enskip ,  \nonumber \\
        \frac{\partial (\rho \ubold)}{\partial t} + \nabla \cdot (\rho \ubold \ubold)
         + \nabla p &=& -\rho g \er \enskip  , \label{orig:mom} \\
        \frac{\partial (\rho E)}{\partial t} + \nabla \cdot (\rho \ubold E + p \ubold) &=&  
        \nabla \cdot (\kappa \nabla T) - \rho g (\ubold \cdot \er) +
        \rho H_{\ext} \enskip , \label{energy_eq_full}
\end{eqnarray}
and an equation of state
\[
 p = p(\rho,T) \enskip .
\]
Here $\rho$, $\ubold$, $T$, and $p$ are the density, velocity,
temperature and pressure, respectively, $E = e + \ubold \cdot \ubold /
2$ is the total energy with $e$ representing the internal energy, and
$g(r)$ is the radially dependent gravitational acceleration (resulting
from spherically symmetric self-gravity), $\er$ is the unit vector in
the radial direction, and $\kappa$ is the thermal conductivity.  The
Reynolds number of flows in a typical white dwarf is sufficiently
large that we neglect viscosity here, though viscous terms could
easily be included in the model and the numerical methodology.

Again we choose to work with enthalpy, $h = e + p/\rho$, rather than
energy, replacing Eq.~[\ref{energy_eq_full}] above by
\begin{equation}
\frac{\partial (\rho h)}{\partial t} + \nabla \cdot (\ubold \rho h)  =
 \frac{Dp}{Dt} + \rho H \enskip , \label{enthalpy_eq} 
\end{equation}
where $\rho H = \rho H_{\ext} + \nabla \cdot (\kappa \nabla T)$
represents the enthalpy source terms, and
$D/Dt = \partial t + \ubold \cdot\nabla$ represents the Lagrangian (or
material) derivative.
(For the purposes of this paper we could alternatively use the entropy equation,
\[
\rho T \frac{D S}{Dt} = \rho H \enskip ,
\]
but for future work in which the source terms due to reactions are
essential we will prefer the enthalpy formulation.) 

We recall from the low Mach number asymptotics in paper~I that 
the assumption that $M = |\ubold| / c \ll 1$ is sufficient to decompose
the pressure into a base state pressure,~$p_0$, and a perturbational
pressure,~$\pi$, i.e. 
\[
p({\x},r,t) = p_0(r,t) + \pi({\x},r,t) \enskip ,
\]
where $\pi / p_0 = O(M^2)$.
Here $\x$ represents the horizontal coordinate directions and $r$ represents
the radial direction.  We define the base state density, $\rho_0$, by assuming
hydrostatic equilibrium of the base state;
this allows us to rewrite Eq.~[\ref{orig:mom}] in the form 
\[
        \frac{\partial (\rho \ubold)}{\partial t} + \nabla \cdot (\rho \ubold \ubold)
         + \nabla \pi = -(\rho  - \rho_0) g \er
\]
with no loss of generality.

Continuing to follow the derivations of paper~I but with a time-dependent base state,
we rewrite conservation of mass as an expression for the
divergence of velocity:
\begin{equation}
\nabla \cdot \ubold = -\frac{1}{\rho} \frac{D \rho}{D t} \enskip .
\label{eq:divu}
\end{equation}

Differentiating the equation of state, $p = p(\rho,T)$, along particle
paths, we can write
\begin{equation}
\frac{D \rho}{Dt}  = \frac{1}{p_\rho}
    \left( \frac{D p}{Dt} - p_T \frac{D T}{Dt} \right) \enskip ,
\label{eq:drdt}
\end{equation}
with $p_\rho = \left.\partial p/\partial \rho\right|_{T}$, 
and $p_T = \left.\partial p/\partial T\right|_{\rho}$.

An expression for $DT/Dt$ can be found by
applying the chain rule to the enthalpy equation (Eq.~[\ref{enthalpy_eq}]):
\begin{equation}
\frac{DT}{Dt} = \frac{1}{\rho c_p} \left( (1 - \rho h_p) \frac{D p}{D t}
 + \rho H \right) \enskip , \label{eq:dTdt}
\end{equation}
where $c_p = \left.\partial h/\partial T\right|_{p}$ is the specific
heat at constant pressure, and $h_p = \left.\partial h/\partial
p\right|_{T}$ for convenience.  Substituting Eq.~[\ref{eq:dTdt}]
into Eq.~[\ref{eq:drdt}] and the resulting expression into
Eq.~[\ref{eq:divu}] yields
\[
\nabla \cdot \ubold = \frac{1}{\rho p_\rho} 
  \left( \frac{p_T}{\rho c_p}(1  - \rho h_p) - 1 \right) \frac{D p}{D t}  
 + \frac{1}{\rho p_\rho} \left(
  \frac{\rho p_T H}{\rho c_p} \right) \enskip ,
\]
still with no loss of generality.
Now, replacing $p$ by $p_0(r,t)$ and recalling the definition of $H$, 
we write the divergence constraint as 
\[
\nabla \cdot \ubold + \alpha \left( \frac{\partial p_0}{\partial t} + \ubold \cdot \nabla p_0 \right) =
\frac{1}{\rho p_\rho} \left(
   \frac{p_T}{\rho c_p} \left(\nabla \cdot (\kappa \nabla T) + \rho H_{\ext} \right)
  \right) \equiv \tilde{S} \enskip ,
\]
where 
\[
\alpha(\rho,T) \equiv - \left( \frac{(1 - \rho h_p )p_T - \rho c_p}{\rho^2
  c_p p_\rho} \right) = \frac{1}{\Gamma_1 p_0} \enskip ,
\]
and $\Gamma_1 \equiv \left. {d (\log p)}/{d (\log \rho)} \right|_s$.
We recall from paper~I that 
$\nabla \cdot \ubold + \alpha \ubold \cdot \nabla p_0 $
can be rewritten as 
$1 / \beta_0 \nabla \cdot (\beta_0 \ubold) $
where 
\begin{equation}
\beta_0(r,t) = \beta(0,t) \exp \left ({\int_0^r \frac{1}{(\Gamma_1 p)_0} 
                          \frac{\partial p_0}{\partial r^\prime} \, dr^\prime} \right ) 
\enskip .
\label{eq:make_beta} 
\end{equation}
Thus we can write the constraint as 
\begin{equation}
 \nabla \cdot (\beta_0 \ubold) = \beta_0 (\tilde{S} - \alpha \frac{\partial p_0}{\partial t})
\enskip .  \label{constraint_eq}
\end{equation}

For the purposes of comparing the fundamental hydrodynamic behavior of this 
low Mach number model to the established compressible
formulation, we will from now on neglect thermal conduction.
Summarizing the low Mach number equation set for this specialized case,
with the momentum equation re-written as an evolution equation for velocity, 
we have 
\begin{eqnarray}
\frac{\partial \rho}{\partial t}  &=& -\nabla \cdot (\rho \ubold)
\enskip , \nonumber \\
\frac{\partial (\rho h)}{\partial t}  &=& -\nabla \cdot (\rho \ubold h) +
\frac{Dp_0}{Dt} + \rho H_{\ext}  \label{eq:enth} \enskip , \\ 
\frac{\partial\ubold}{\partial t} &=& -\ubold \cdot \nabla \ubold
 -\frac{1}{\rho} \nabla\pi - \frac{(\rho - \rhozero)}{\rho} g \er
 \enskip , \nonumber \\ 
\nabla \cdot (\beta_0 \ubold )&=& \beta_0 \left( 
\sigma
H_{\ext} - \frac{1}{\Gamma_1 p_0} \frac{\partial p_0}{\partial t} \right) \label{eq:constr} \enskip ,
\end{eqnarray}
where we define, for convenience, $\sigma = p_T / (\rho c_p p_\rho)$. 

This system differs from that in paper~I in
that now $p_0$ and $\rhozero$ are unknowns as well as $\rho$, $\rho h$, $\ubold$,
and $\pi$. The equation of
state was used to derive the constraint thus to include it here would
be redundant.  When reactions and compositional effects are included
in future work, evolution equations for species will be added to
this system and reaction terms will be added to the enthalpy equation
and divergence constraint, but for the hydrodynamical tests we present
here this system is sufficient.

We follow the approach used in \citet{almgren:2000} to compute
the time evolution of the base state, recalling from the beginning of this section
that the pressure of each parcel remains unchanged during base state adjustment, i.e.,
\begin{equation}
\frac{D p_0}{D t} = 0 \enskip . \label{eq:dpdtzero}
\end{equation}
We first calculate the radial velocity field, denoted $w_0$, 
that adjusts the base state.  We decompose the full velocity field, $\ubold$, into $w_0 \er$
and the remaining velocity field, $\ut$, that governs the more local dynamics,
i.e.,
\begin{equation}
\ubold(\x,r,t) = w_0(r,t) \; \er +   \ut(\x,r,t) \enskip , \label{eq:vel_decomp}
\end{equation}
and write Eq.~[\ref{eq:constr}] in terms of 
$w_0$ and $\ut$,  
\begin{eqnarray}
\nabla \cdot (\beta_0 w_0 \er) + \nabla \cdot (\beta_0 \tilde{\ubold}) &=& \beta_0 \left( 
\sigma
H_{\ext} - \frac{1}{\Gamma_1 p_0} \frac{\partial p_0}{\partial t} \right) \label{eq:constr_decomp}\enskip .
\label{eq:jbb1}
\end{eqnarray}
We then integrate Eq.~[\ref{eq:jbb1}] over a horizontal slab $\Omega_H \times (r-h, r+h)$ to obtain
\begin{equation}
\int_{r-h}^{r+h}
\int_{\Omega_H} \left( \nabla \cdot (\beta_0 w_0 \er) + \nabla \cdot (\beta_0 \tilde{\ubold}) \right)  dr \;  d\x
= \int _{r-h}^{r+h}
\int_{\Omega_H} \beta_0 \left( \sigma
H_{\ext} - \frac{1}{\Gamma_1 p_0} \frac{\partial p_0}{\partial t} \right)  dr \;  d\x ,
\end{equation}
Assuming solid wall or periodic boundary conditions on the horizontal boundaries, 
or that the horizontal velocity decays sufficiently as we reach the horizontal
boundaries,  we can simplify the volume integrals 
into area integrals over $\Omega_H,$
\begin{eqnarray}
\left .
\int_{\Omega_H} \left( (\beta_0 w_0) + (\beta_0 \tilde{\ubold}) \cdot \er \right) d\x  \right |_{r-h}^{r+h}
&=&
\int_{r-h}^{r+h}
\int_{\Omega_H} \beta_0 \left( \sigma
H_{\ext} - \frac{1}{\Gamma_1 p_0} \frac{\partial p_0}{\partial t} \right) dr \;  d\x ,
\label{eq:jbb2}
\end{eqnarray}
To define the model we want to partition the velocity field so that
the horizontal average of the radial flux due to heating be entirely
incorporated into $w_0$ rather than $\ut$, namely
\[
\int_{\Omega_H}  \ut \cdot \er \; d\x \equiv 0 \enskip , 
\]
Using this relationship, and taking the limit as $h\rightarrow 0$ Eq.~[\ref{eq:jbb2}] can be simplified to 
\begin{eqnarray}
\frac{\partial (\beta_0 w_0)}{\partial r}
&=&  \beta_0 \left( (\overline{\sigma H}) - \frac{1}{\Gamma_1 p_0}
  \frac{\partial p_0}{\partial t}  \right) \enskip .
\label{eq:divbeta0w0}
\end{eqnarray}
where we define $\overline{\sigma H} = 1 / {\rm Area}(\Omega_H) \int_{\Omega_H} ( \sigma H_{\ext} ) \; d\x$.

We can further simplify Eq.~[\ref{eq:divbeta0w0}] by
expanding $\partial(\beta_0 w_0) / \partial r = \beta_0 {\partial     w_0}/{\partial r} 
                                           +    w_0 {\partial \beta_0}/{\partial r}$, 
exploiting $D p_0 / D t = {\partial p_0}/{\partial t} + w_0 {\partial p_0}/{\partial r} = 0$ 
to replace ${\partial p_0}/{\partial t}$, and recalling from the definition
of $\beta_0$ that $(1/\beta_0) \; {\partial \beta_0}/{\partial r} = 
(1/\Gamma_1 p_0) \; {\partial p_0}/{\partial r}$ (see Appendix~B of paper~I
for the derivation of $\beta_0$.)
Then
\[
\frac{\partial     w_0}{\partial r}  = \overline{\sigma H}  \enskip .
\]
This system can be integrated by noting that if there exists a lower
boundary at $r=r_0$ with zero normal velocity (such as the center of a star), then
at any time, $t,$ 
\begin{equation}
w_0(r,t) = \int_{r_0}^r \overline{\sigma H} (\rp,t) \; d \rp \enskip . \label{eq:w0int}
\end{equation}
The base state pressure and density update follow from Eq.~[\ref{eq:dpdtzero}]
and conservation of mass, respectively:
\begin{eqnarray}
\frac{\partial      p_0}{\partial t} &=& - w_0 \frac{\partial p_0}{\partial r}   \label{eq:p0upd} \\
\frac{\partial \rhozero}{\partial t} &=& -\frac{\partial (\rhozero w_0)}{\partial r} \label{eq:rho0upd} \enskip .
\end{eqnarray}
There are now two choices for defining the new base state enthalpy; these
options are analytically equivalent but may differ numerically.
The first is to use the equation of state: $(\rho h)_0 = \rho_0 h(p_0, \rho_0)$.
The second is to use Eq.~[\ref{eq:enth}].   In this second approach, we 
can exploit $D p_0 / D t = 0 $, but must also correctly partition
the heating term in the right hand side of Eq.~[\ref{eq:enth}]. The partitioning
is given by the requirement that the base state continue to satisfy the
equation of state.  Then, given $D p_0 / Dt = 0$,
\begin{eqnarray*}
\frac{D h_0}{D t} &=& \left.\frac{\partial h_0}{\partial T   }\right|_p \frac{D T_0}{D t} 
                 \;  = \;  c_p (-\frac{p_\rho}{p_T}) \frac{D \rho_0}{D t}  
                 \;  = \;  \frac{\rho_0 c_p p_\rho}{p_T} \overline{\sigma H} \\
                  &=&  \frac{1}{\sigma_0} \overline{\sigma H} \enskip ,
\end{eqnarray*}
recalling $\sigma = p_T / (\rho c_p p_\rho)$ and letting $\sigma_0 = \sigma(p_0, \rho_0)$.
Returning to conservation form, we can write
\begin{equation}
\frac{\partial (\rho h)_0}{\partial t} = -\frac{\partial (w_0 (\rho h)_0)}{\partial r}
+ \frac{\rhozero}{\sigma_0} \; \overline{\sigma H} \enskip .
\label{eq:rhoh0upd} 
\end{equation}

Using the velocity decomposition (Eq.~[\ref{eq:vel_decomp}]) we can re-write the evolution equations
for $\rho$ and $\rho h$ as 
\begin{eqnarray}
\frac{\partial \rho}{\partial t}  &=& -\nabla \cdot (\rho \ut) - \frac{\partial (\rho w_0)}{\partial r}  
\label{eq:rhotildeupd} \\
\frac{\partial (\rho h)}{\partial t}  &=& -\nabla \cdot (\rho h \ut) 
  - \frac{\partial (\rho h w_0)}{\partial r} 
  + \tilde{w} \frac{\partial p_0}{\partial r} 
  + \rho H_{\ext} \enskip , \label{eq:rhohtildeupd} 
\end{eqnarray}
where $\tilde{w} = \ut \cdot \er$.
We can also write these in perturbational form (with no loss of generality):
\begin{eqnarray*}
\frac{\partial \rho^\prime}{\partial t}  
&=& -\nabla \cdot (\rho^\prime (\ut + w_0 \er)) -\nabla \cdot (\rhozero \ut)
\\ 
\frac{\partial (\rho h)^\prime}{\partial t}  &=& 
 -\nabla \cdot ((\rho h)^\prime (\ut + w_0 \er)) -\nabla \cdot ((\rho h)_0 \ut)
  + \tilde{w} \frac{\partial p_0}{\partial r} 
+ \left( \rho H_{\ext} - \frac{\rhozero}{\sigma_0} \overline{\sigma H} \right) \enskip , 
\end{eqnarray*}
using Eq.~[\ref{eq:rho0upd}] and Eq.~[\ref{eq:rhoh0upd}],
where $\rho^\prime \equiv \rho - \rhozero$ and $(\rho h)^\prime \equiv (\rho h) - (\rho h)_0$.

The evolution of the velocity field becomes
\begin{equation}
\frac{\partial\ut}{\partial t} = -\ut \cdot \nabla \ut - w_0 \frac{\partial \ut}{\partial r}
                                   -\tilde{w} \frac{\partial w_0}{\partial r} \er
 -\frac{1}{\rho} \nabla\pi - \frac{(\rho-\rhozero)}{\rho} g \er  \label{eq:utildeupd} \enskip ,
\end{equation}
and subtracting Eq.~[\ref{eq:divbeta0w0}] from Eq.~[\ref{eq:constr}],  
the constraint equation for $\ut$ becomes
\begin{equation}
\nabla \cdot (\beta_0 \ut )  = \beta_0 (\sigma H)^\prime \label{eq:tildeconstraint} \enskip .
\end{equation}
where we define $(\sigma H)^\prime = \sigma H_{\ext} - (\overline{\sigma H})$.

In summary, then, the evolution of the base state is described by 
Eq.~[\ref{eq:p0upd}] and Eq.~[\ref{eq:rho0upd}],
with $w_0$ given by Eq.~[\ref{eq:w0int}], and the evolution of the full state is given by 
Eq.~[\ref{eq:rhotildeupd}], Eq.~[\ref{eq:rhohtildeupd}] and Eq.~[\ref{eq:utildeupd}] with the
divergence constraint given by Eq.~[\ref{eq:tildeconstraint}].

\section{Numerical Methodology}
\label{sec:numer}

Our strategy for evolving the low Mach number system with a time-varying
base state is a fractional step approach. In each time step 
we first update density and enthalpy
as if the base state were time-independent, giving us predicted
values that can be used to construct time-centered values in the right hand side 
of Eq.~[\ref{eq:w0int}]. We then compute the evolution of the
base state and recompute the updates to density and enthalpy, incorporating 
the base state adjustment. Finally, we
update and project the velocity field to define the new values of
velocity and pressure.  The upwind methodology used to update all
the state variables provides a robust discretization of the
convective terms that avoids any stability restriction other than
the CFL constraint, i.e. the time step scales linearly with the
grid spacing and inversely with the maximum magnitude of the velocity 
in any one coordinate direction in the domain.

All base state quantities as well as all state quantities other than
the perturbational pressure, $\pi$, are defined at cell centers and
integer time levels.  The perturbational pressure is defined at nodes 
and at half-times; similarly, the advective velocity and fluxes used for 
advective updates are defined at edges and half-times.  In this section we replace $\ut$
by $\ubold$ for convenience of notation.

\paragraph{Initialization}  
Specification of the initial value problem includes initial values for
$p_0, \rho_0$ and $h_0$ (or $T_0$) as well as $U, \rho$ and $h$ (or $T$)
at time $t=0,$ and a description of the boundary
conditions,  but the perturbational pressure is not initially prescribed.
We calculate $\beta_0$ at $t=0$ using Eq.~[\ref{eq:make_beta}].  
Given this initial $\beta_0$, we project the initial 
velocity field to ensure that it satisfies the divergence constraint at $t=0$.
Then initial iterations of the following steps (typically two are sufficient) are 
performed to calculate an approximation to the perturbational pressure at $t = {\Delta t}/{2}$.
At the end of each initial iteration all variables other than $\pi$ are
reset to their initial values.

The following steps are components of the single time step taken to advance
the solution from $t^n$ to $t^{n+1}.$

\paragraph{Step 1}  In this step we construct $\uadv,$ a 
time-centered, second-order accurate, staggered-grid approximation to $\ubold$ at $t^{\nph},$
using an unsplit second-order Godunov procedure \citep{colella1990}.
To do so we first predict $\uadvstar$ using the
cell-centered data at $t^n$ and the lagged pressure gradient
from the interval centered at $t^\nmh$.
The provisional field, $\uadvstar$, represents a normal velocity on cell edges
analogous to a MAC-type staggered grid discretization of the
Navier-Stokes equations \citep{harlowwelch}.  However,
$\uadvstar$ fails to satisfy the time-centered divergence
constraint (Eq.~[\ref{eq:tildeconstraint}]).
We apply a discrete projection by solving the elliptic equation
\[
D^\mathrm{MAC} (\frac{\beta_0^n}{\rho^n}G^\mathrm{MAC} \phi^\mathrm{MAC}) =
D^\mathrm{MAC} (\beta_0^n \uadvstar) - \beta_0^n \left( (\sigma H)^\prime \right)^n
\]
for $\phi^\mathrm{MAC}$, where $D^\mathrm{MAC}$ represents a centered
approximation to a cell-based divergence from edge-based velocities,
and $G^\mathrm{MAC}$ represents a centered approximation to edge-based
gradients from cell-centered data.  The solution, $\phi^\mathrm{MAC}$, is
then used to define
\[ 
\uadv = \uadvstar - \frac{1}{\rho^n} G^\mathrm{MAC} \phi^\mathrm{MAC} \enskip .
\]
In the above equations, we average $\beta_0^n$ and $\rho_0^n$ 
from centers to edges, i.e., 
${\beta_0}_{j+\myhalf}^n = 1/2 ({\beta_0}_{j}^n + {\beta_0}_{j+1}^n)$, 
${\rho}_{i+\myhalf,j}^n = 1/2 ({\rho}_{i,j}^n + {\rho}_{i+1,j}^n)$ and 
${\rho}_{i,j+\myhalf}^n = 1/2 ({\rho}_{i,j}^n + {\rho}_{i,j+1}^n)$.

\paragraph{Step 2}
We update $\rho$ and $\rho h$ as if $w_0 = 0$ and the base state were constant, 
i.e., we discretize
\begin{eqnarray*}
\frac{\partial \rho}{\partial t}  &=& - \nabla \cdot (\rho^\prime \ubold) 
                                      - \nabla \cdot (\rhozero    \ubold) \enskip , \\
\frac{\partial (\rho h)}{\partial t}  &=& -\nabla \cdot ((\rho h)^\prime \ubold) 
                                          -\nabla \cdot ((\rho h)_0      \ubold)  
                                          + w \frac{\partial p_0}{\partial r}
                                          + \rho H_{\ext}
\end{eqnarray*}
using the second-order advection methodology as in paper~I with 
$\rho H_{\ext}$ treated as an explicit source term.  
The discretization takes the form 
\begin{eqnarray*}
\rho^{n+1,*} &=& \rho^n - \dt \left[ \nabla \cdot (\rho^\prime \uadv) \right]^{\nph} 
                        - \dt        \nabla \cdot (\rhozero^n  \uadv)        
\enskip , \\
(\rho h)^{n+1,*} &=& (\rho h)^n - \dt \left[ \nabla \cdot ((\rho h)^\prime \uadv) \right]^{\nph} 
                                - \dt        \nabla \cdot ({(\rho h)_0}^n  \uadv)         \\
                 &&             + \dt \; \wadv  \left(\frac{\partial p_0}{\partial r} \right)^n
                                + \dt \left( \rho H_{\ext} \right)^{\nph} \enskip ,
\end{eqnarray*}
where $\wadv = \uadv \cdot \er$ and  $\rho^{\nph} = 1 / 2 (\rho^n + \rho^{n+1,*})$
in the construction of $\left( \rho H_{\ext} \right)^{\nph}$.
The details of the upwind construction of 
$\left[ \nabla \cdot (\uadv \rho^\prime) \right]^{\nph}$
and $\left[ \nabla \cdot (\uadv (\rho h)^\prime) \right]^{\nph}$
are given in Appendix~A, where we consider the construction
of $\left[ \nabla \cdot (\V s) \right]^{\nph}$ for any edge-based 
vector field $\V$ and cell-centered quantity $s$.  In this step $\V = \uadv$.
The terms, $\nabla \cdot (\rho_0^n      \uadv) $ and
           $\nabla \cdot ((\rho h)_0^n      \uadv)$,
are defined differently, in that we do not upwind $\rho_0$ or $(\rho h)_0$ in this step, 
rather they are simply averaged onto edges as $\beta_0$ and $\rho^n$ were averaged
in \textbf{Step 1}.

\paragraph{Step 3}
We integrate Eq.~[\ref{eq:w0int}] to determine $w_0$
on edges, 
\[
{w_0}_{j+\myhalf} = {w_0}_{j-\myhalf} + \dr (\overline{\sigma H})_j^{\nph} \enskip ,
\]
using the equation of state given $\rho^{\nph}$ 
and $p_0$ to compute $\sigma$. We then update the base state quantities, 
\begin{eqnarray*}
(p_0)_j^{n+1} &=& (p_0)_j^n - \frac{\dt}{2 \dr} 
( {w_0}_{j+\myhalf} + {w_0}_{j-\myhalf} )
( {p_0}_{j+\myhalf}^{\nph} -  {p_0}_{j-\myhalf}^{\nph} )
\enskip , \\
(\rho_0)_j^{n+1} &=& (\rho_0)_j^n - \frac{\dt}{\dr} 
( (\rho_0 w_0)_{j+\myhalf}^{\nph} -  (\rho_0 w_0)_{j-\myhalf}^{\nph} ) \enskip , \\
(\rho h)_0^{n+1} &=& \rho_0^{n+1} \; h(p_0^{n+1}, \rho_0^{n+1}) \enskip .
\end{eqnarray*} 
The construction of ${p_0}_{j+\myhalf}^{\nph}$ and ${\rho_0}_{j+\myhalf}^{\nph}$
is described in Appendix~A.
After construction of the new base state we compute $\beta_0^{n+1}$ using
Eq.~[\ref{eq:make_beta}],  then 
set $\beta_0^{\nph} = 1 / 2 (\beta_0^n + \beta_0^{n+1})$.
We use the equation of state here to calculate $(\rho h)_0^{n+1}$ in order to
keep the base state thermodynamically consistent.

\paragraph{Step 4}
In this step we repeat the update of $\rho$ and $\rho h$, but in the
prediction of the edge states here $\V = \uadv + w_0 \er$.
We also center the $w \; \partial p_0 / \partial r$ term in time:
\begin{eqnarray*}
\rho^{n+1} &=& \rhozero^{n+1} + (\rho^n - \rhozero^n) 
 - \dt \left[ \nabla \cdot (\rho^\prime (\uadv+w_0 \er)) \right]^\nph  
 - \dt        \nabla \cdot (\rhozero^n \uadv)    \\
(\rho h)^{n+1} &=& (\rho h)_0^{n+1} + ((\rho h)^n - (\rho h)_0^n) 
 - \dt \left[ \nabla \cdot ((\rho h)^\prime (\uadv+w_0 \er)) \right]^\nph  \\
&& - \dt        \nabla \cdot ((\rho h)_0^n \uadv)  \\
&& + \frac{\dt}{2} \; \wadv \left( \left(\frac{\partial p_0}{\partial r} \right)^{n+1} 
  + \left(\frac{\partial p_0}{\partial r} \right)^{n}  \right)
 + \left( \rho H_{\ext} - \frac{\rhozero}{\sigma_0} \overline{\sigma H} \right)^{\nph} 
\enskip .
\end{eqnarray*}
We use the perturbational form of these equations in order to ensure that numerically,
if, as in the anelastic case, $\beta_0 \equiv \rho_0$, and $\rho^n = \rhozero^n$ and $H$ is 
horizontally uniform, then $\rho^{n+1} = \rhozero^{n+1}$, i.e. no perturbation to the
base state density is introduced in a case where analytically there should be none.
 
\paragraph{Step 5}
We then update the velocity field, $\ubold^n$ to $\ubold^{n+1,*}$ by discretizing 
Eq.~[\ref{eq:utildeupd}],
\begin{eqnarray*}
\ubold^{n+1,*} &=& \ubold^n - \dt  \left[((\uadv+w_0 \er) \cdot \nabla) \ubold\right]^\nph 
      - \dt w^{\mathrm{ADV}}  \left(\frac{\partial w_0^{\nph}}{\partial r} \right) \er \\
   &&   - \frac{\dt}{\rho^\nph} \mathbf{G} \pi^\nmh 
        - \dt \frac{(\rho^\nph-\rhozero^\nph)}{\rho^\nph} g \er \enskip ,
\end{eqnarray*}
with $\rho^\nph = \myhalf ( \rho^n + \rho^{n+1} )$ 
and $\mathbf{G}$ a discretization of the gradient operator.
The construction of $\left[((\uadv+w_0 \er) \cdot \nabla) \ubold\right]^\nph$ is
described in Appendix~A with $\V = \uadv+w_0 \er$ and~$s$ set to each
component of $\ubold^n$ individually.
Finally, we impose the constraint (Eq.~[\ref{eq:tildeconstraint}])
\[
\nabla \cdot (\beta_0^{\nph} \ubold^{n+1} )  = \beta_0^{\nph} \left({(\sigma H)^\prime}\right)^{n+1} 
\]
by solving 
\[
 L_\beta^\rho \phi =
   D \left( \beta_0^{\nph} (\frac{\ubold^{n+1,*}}{\Delta t} + \frac{1}{\rho^\nph} \mathbf{G} \pi^\nmh) \right)
- \frac{\beta_0^{\nph} \left( (\sigma H)^\prime\right)^{n+1}}{\Delta t}
\]
for nodal values of $\phi$ where
$L_\beta^\rho$ is the standard bilinear
finite element approximation to $\nabla \cdot ({\beta_0}/{\rho}) \nabla$
with $\rho$ and $\beta_0$ evaluated at $t^\nph$.  
(See \citet{almgrenBellSzymczak:1996} for a detailed discussion of this
approximate projection; see \citet{almgren:bell:crutchfield} for a discussion
of this particular form of the projection operand.)
We determine the new-time velocity field from
\[
\ubold^{n+1} = \ubold^{n+1,*} - \frac{\Delta t}{\rho^\nph}
          \left( \mathbf{G} \phi - \mathbf{G} \pi^\nmh \right) \enskip ,
\]
and the new time-centered perturbational pressure from
\[
  \pi^\nph = \phi \enskip .
\]

\section{Numerical Results}
\label{sec:results}

We consider three numerical tests in this section, each studying the response of
the atmosphere to prescribed external heating.  
For each test case, the initial conditions in the computational
domain are specified in two parts.  The lower portion of the domain
is initialized with a one-dimensional hydrostatic white dwarf model
up until the outer boundary of the white dwarf.  This initialization is
identical for the compressible and low Mach number models.
The model is created by specifying a base density of $2.6\times 10^9~\gcc$ and base
temperature of $7\times 10^8$~K and integrating the equation of
hydrostatic equilibrium outward while constraining the model to be
isentropic.  The composition is held constant at 0.3~$^{12}$C and
0.7~$^{16}$O, and the gravitational acceleration is fixed at $-1.5\times
10^{10}$~cm~s$^{-2}$.  We use the stellar equation of state developed
by \cite{timmes_swesty:2000}.  This procedure provides a reasonable
approximation of the state of the white dwarf just before runaway.
None of the methods described here require constant gravity, but it
was assumed for simplicity in the comparisons.  

\begin{figure*}
\begin{center}
\plotone{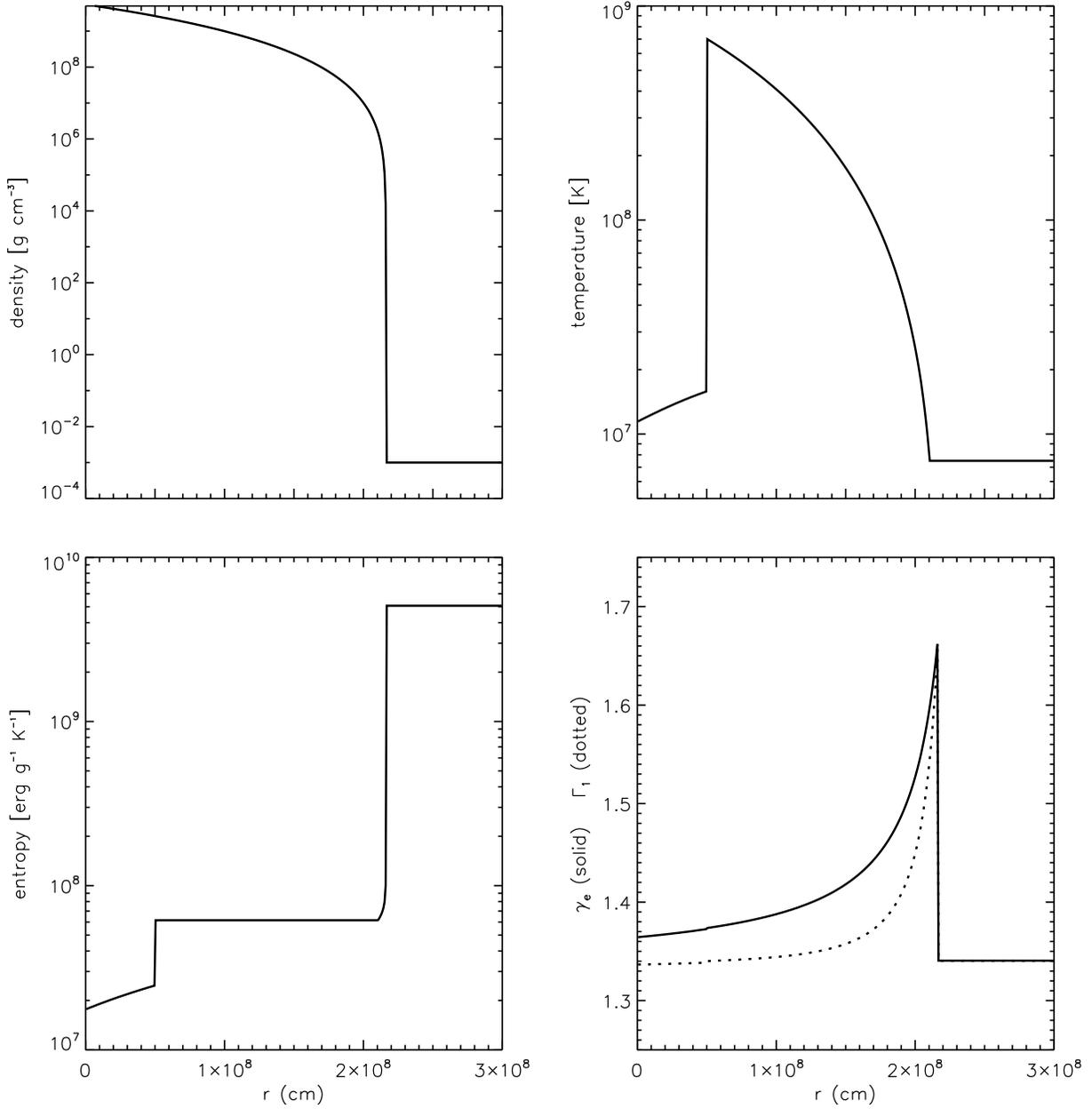}
\end{center}
\caption{\label{fig:initialmodel} The white dwarf atmosphere initial
model.  Shown are the density (top left), temperature (top right),
entropy (bottom left), and adiabatic indices (bottom right).  To
prevent convective motions from hitting the lower boundary in our
multidimensional tests, the first $5\times 10^7$~cm of this model is
constructed to have a convectively stable entropy profile.  The
one-dimensional tests do not use this portion of the model.}
\end{figure*}

The upper portion of the domain represents the region beyond the outer boundary
of the white dwarf, and different approximations are used there for
the compressible and low Mach number models.
For the compressible calculations, the integration proceeds radially
outward until the density reaches a threshold value of $10^{-4}~\gcc$.
Throughout the integration we set a low temperature cutoff of
$10^7$~K, to keep the temperature in the outer layers of the model
reasonable.  Once the density drops below its cutoff, the integration
is stopped and the material above it is held at constant density and
temperature.  This buffer region is necessary to allow for expansion
of the star; otherwise, as the star expands the loss of mass through
the upper domain boundary would change the base pressure
\citep{glasner:2005}, impacting the dynamics throughout the domain.
Finally, for the multidimensional test cases, we add a
convectively stable layer below the atmosphere to prevent any motions
generated from the heating from interfering with the lower boundary.
Figure~\ref{fig:initialmodel} shows the initial temperature, density,
entropy, and adiabatic indices ($\Gamma_1$ and $\gamma_e \equiv
p/(\rho e) + 1$) as a function of height for the compressible
background.  

For the low Mach number model applied to the second and third test cases,
the density cutoff is set to $2.5 \times 10^{6}~\gcc$, approximately the value at 
which the temperature cutoff is applied for the compressible background.
Once the density reaches this cutoff the density, temperature and pressure
are held constant, equivalent to gravity being set to zero radially 
outward of that position.  Because the base state density in the buffer region
is significantly higher for the low Mach number calculations than for
the compressible calculations, this buffer region serves to damp
motions that reach it without impacting the hydrostatic equilibrium of
regions toward the center.  Since the time step for the entire calculation
is determined by the largest velocity in the domain,  
this damping is essential for low Mach number calculations 
in order to avoid excessively large velocities above the cutoff 
that would dictate an excessively small time step.
An additional approximation in the outer region is that we set
$\beta_0 \equiv \rho_0$ for $\rho_0 < 5 \times 10^7~\gcc$ in order to suppress spurious
wave formation at the outer boundary of the star. 

In the first test, a layer of the star is heated for 5.0~s with a heating profile, 
\[
H = H_0 \exp(-(r-r_0)^2 / W^2) \enskip ,
\]
with $r_0 = 4\times 10^7$~cm, $W = 10^7$~cm, and $H_0 = 1\times
10^{17}$~erg~g$^{-1}$~s$^{-1}$.  This energy generation
rate is quite a bit higher than we would expect during the
smoldering phase of the convection leading up to an SN~Ia
\citep{Woosley:2004}, but necessary to see a response with the
compressible code on a reasonable timescale.  It also provides a more
stringent test of the hydrostatic adjustment than a lower energy
generation rate would give.  Because the initial conditions and the
heating are both one-dimensional, we use a reduced one-dimensional
form of the equations to solve the systems.  We contrast three
different systems of equations.  

For this one-dimensional test, we compare the low Mach number results
to those produced by the fully compressible PPM method \citep{ppm}, as
implemented in the FLASH code \citep{flash}.  Two versions of the low Mach number
algorithm are used for this test.  The first is the low
Mach number equation set with time-varying base state (as described in
\S~\ref{sec:lmn}).  The second is  a formulation
of the low Mach number equations in which the base state is
time-independent (equivalent to the equation set present in paper~I).

\begin{figure*}

\begin{center}
\epsscale{0.7}
\plotone{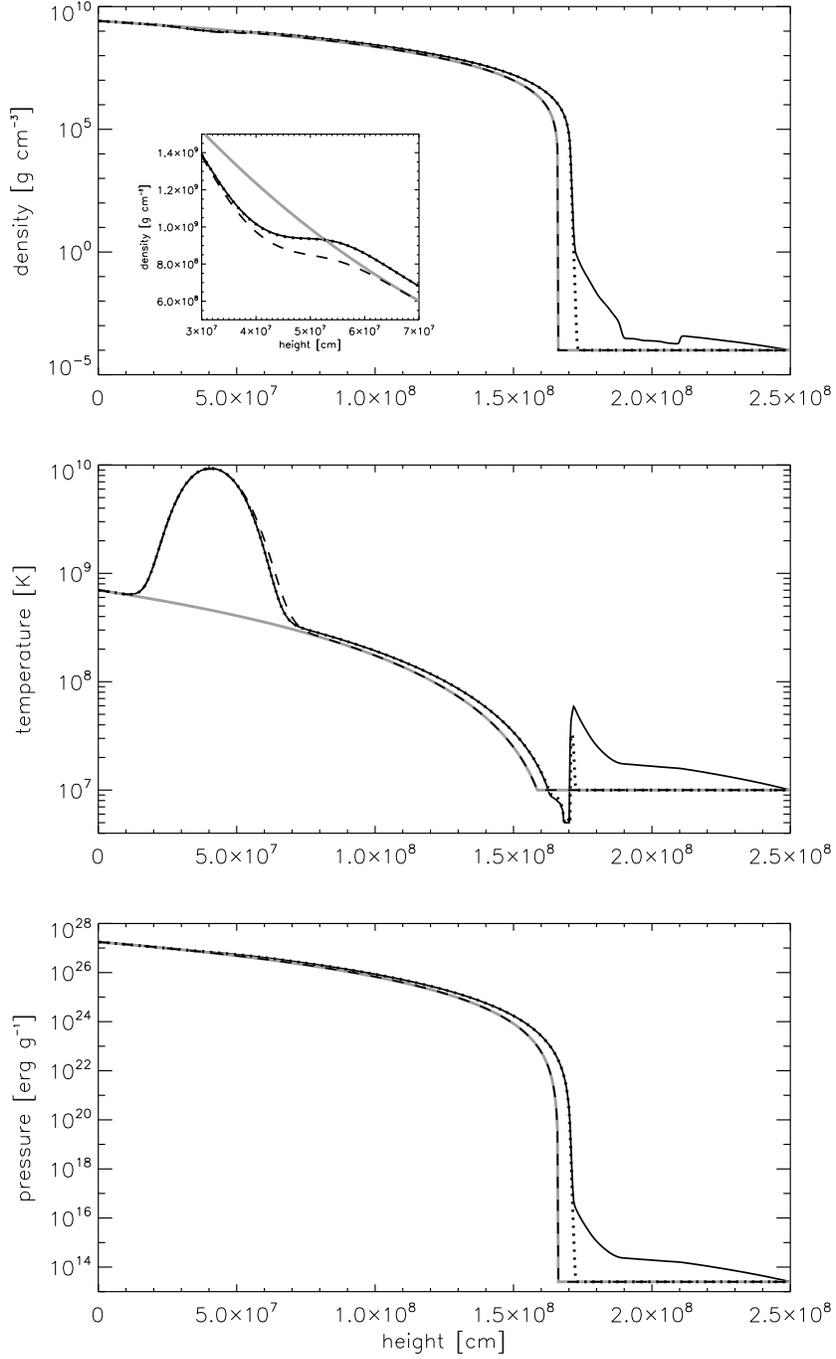}
\epsscale{1.0}
\end{center}

\caption{Hydrostatic adjustment problem with uniform heating:
density (top), temperature (middle) and pressure (bottom).  
The initial conditions are shown in gray.  
The solid black line is the fully compressible solution,  
the dotted line is the low Mach number formulation that allows for the base state expansion,
and the dashed line is the low Mach number formulation assuming a fixed base state.  
The compressible and expanding base state low Mach number solutions 
show excellent agreement.  The low Mach number model with a fixed base state is unable to capture
the correct solution.  The inset in the density plot shows the structure in the vicinity of the local heating.}
\label{fig:oned_compare}
\end{figure*}

Figure~\ref{fig:oned_compare} shows the density, temperature and
pressure for the three solutions at $t=5$~s.  All simulations were run
on a uniform grid with 768 zones spanning $2.5 \times 10^{8}$cm.
The compressible and low Mach number expanding background solutions show excellent
agreement.  In the region of heating, the temperature has
increased enormously, with a corresponding density decrease.  The
amplitude of the density decrease is much smaller, due to the
degenerate nature of the equation of state.  The inset in the density
plot shows the density adjustment on a linear scale, showing the
decrease at the heating height and an increase in the density above
this.  Both equation sets reach the same solution in response to
the heating.  The differences above $1.7\times 10^8$~cm are due to the
different treatments of the upper boundary, and are not significant to
the atmospheric dynamics.  As a result of the energy deposition, the
model expands by almost $10^7$~cm, or about 5\%.  In contrast, when
the background is not allowed to adjust (dashed line), the low Mach
number model fails to capture the expansion.  While the temperature
increases in the region of the heating, there is no expansion in the
material above the heating layer.  This is consistent with the point
made in \cite{bannon:1996a} that the pseudo-incompressible equation
set does not give the correct solution in the terrestrial atmosphere
when the base state is not allowed to vary in time, and the
demonstration in \cite{almgren:2000} that the correct solution is
found when the base state does absorb the horizontally averaged
heating.

\begin{figure*}

\begin{center}
\epsscale{0.7}
\plotone{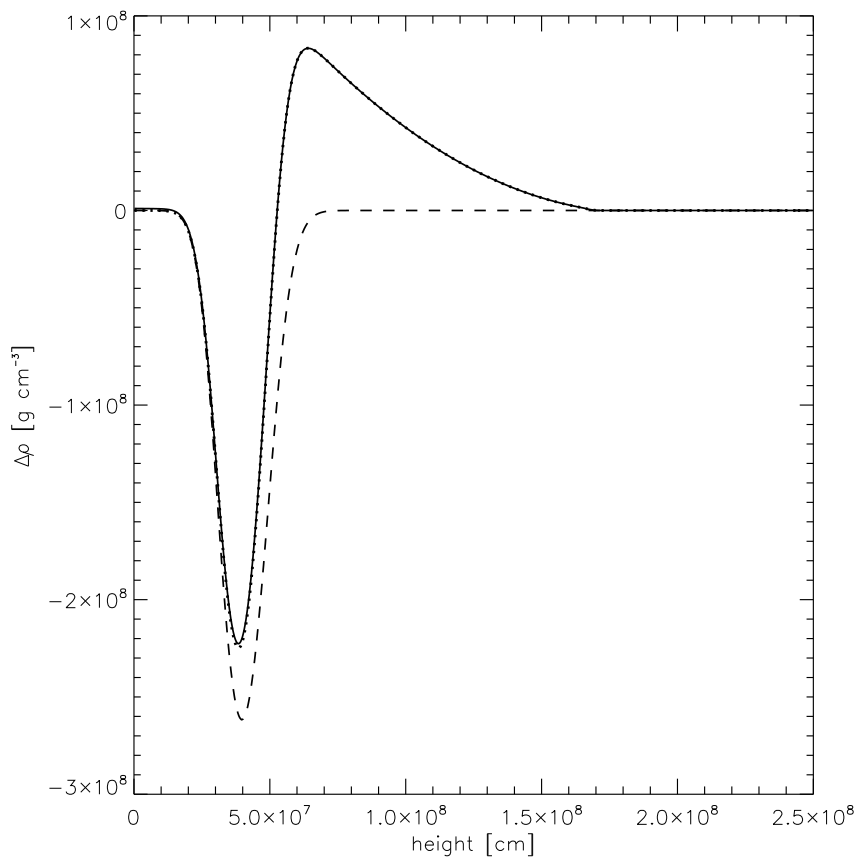}
\epsscale{1.0}
\end{center}

\caption{Hydrostatic adjustment problem with uniform heating: 
the difference in density between $t=0$ and $t=5$~s.
The solid black line is the fully compressible solution,  
the dotted line is the low Mach number formulation that allows for the base state expansion,
and the dashed line is the low Mach number formulation assuming a fixed base state.  
We see close agreement between the compressible and low Mach number formulation
with the time-dependent base state.}
\label{fig:oned_compare_diff}
\end{figure*}

Figure~\ref{fig:oned_compare_diff} shows the difference in density
before and after heating.  Here we see more clearly the adjustment of
the density structure as a result of the localized heating.  There is
a slight rise in the density in the compressible solution below the
heated layer.  This small error is due to the difficulty in constructing a
lower hydrostatic boundary that allows sound waves to leave the
domain.  For the low Mach number case, the state remains unchanged
below the heating layer, as it should.
The compressible and expanding background low Mach
number solutions show a large increase in the density above the
heating layer.  This is not present in the low Mach number case where
the background was not allowed to expand.

In even this one-dimensional simulation, the choice of boundary
conditions at the top and bottom boundaries of the computational
domain is critical for the compressible calculations.
The fully compressible code generates sound waves as it 
struggles to keep the hydrostatic solution steady on the grid.  
The boundary conditions must allow these disturbances to
leave the domain or they will corrupt the solution.  We use a
hydrostatic lower boundary, integrating the pressure and density in
the ghost cells using a fourth-order reconstruction of the pressure
\citep{ppm-hse}, where the temperature is kept constant in the
ghost cells and the velocities are given a zero gradient.  This
provides pressure support to the material while allowing sound waves to
leave the domain.  Further robustness is obtained by computing the
hydrostatic structure in the boundary from the initial base density
and temperature, and keeping this structure fixed in time.  At the
upper boundary, we set the density and temperature to our cutoff
values and allow the fluid to move out of the domain (with a
zero gradient) but set inward velocities to zero.  This has the
effect of keeping the fluid velocities in this region small, and
therefore, they do not dictate the time step.

The low Mach number calculations, by contrast, are not sensitive to
mass exiting the top boundary, as the hydrostatic equilibrium is
incorporated into the base state, which is independent of the total
mass.  A simple outflow boundary condition is used at the top
(with any inflowing velocities set to zero), and a reflecting boundary 
condition is used at the bottom boundary.
Additionally, the compressible calculations need to resolve the
scale height of the atmosphere very well to suppress any ambient
velocities generated by slight imbalances of the pressure gradient and
gravitational force (see \citealt{ppm-hse} for a discussion of this).  This is not the case with the low Mach number
method, so we expect our low Mach number solutions to numerically converge 
at a lower resolution than the compressible solutions.  

For the second and third tests, we consider fully two-dimensional
heating profiles, in which both parts of the new low Mach number
algorithm are fully exercised.  For these cases we compare the new low
Mach number formulations with the compressible solution.  In addition
to the PPM algorithm, these multidimensional tests are also run with an
unsplit compressible algorithm \citep{colella1990}, adapted to a
general EOS, and incorporated into the FLASH framework.  This is the
same implementation of the unsplit method we described in paper I.
The computational domain for these tests is $2.5 \times 10^{8}$cm
by $3.5\times 10^{8}$cm, spanned by a uniform grid with 640$\times$ 896 zones.
Periodic boundary conditions are used on the sides of the domain.

\begin{figure*}

\begin{center}
\epsscale{0.75}
\plotone{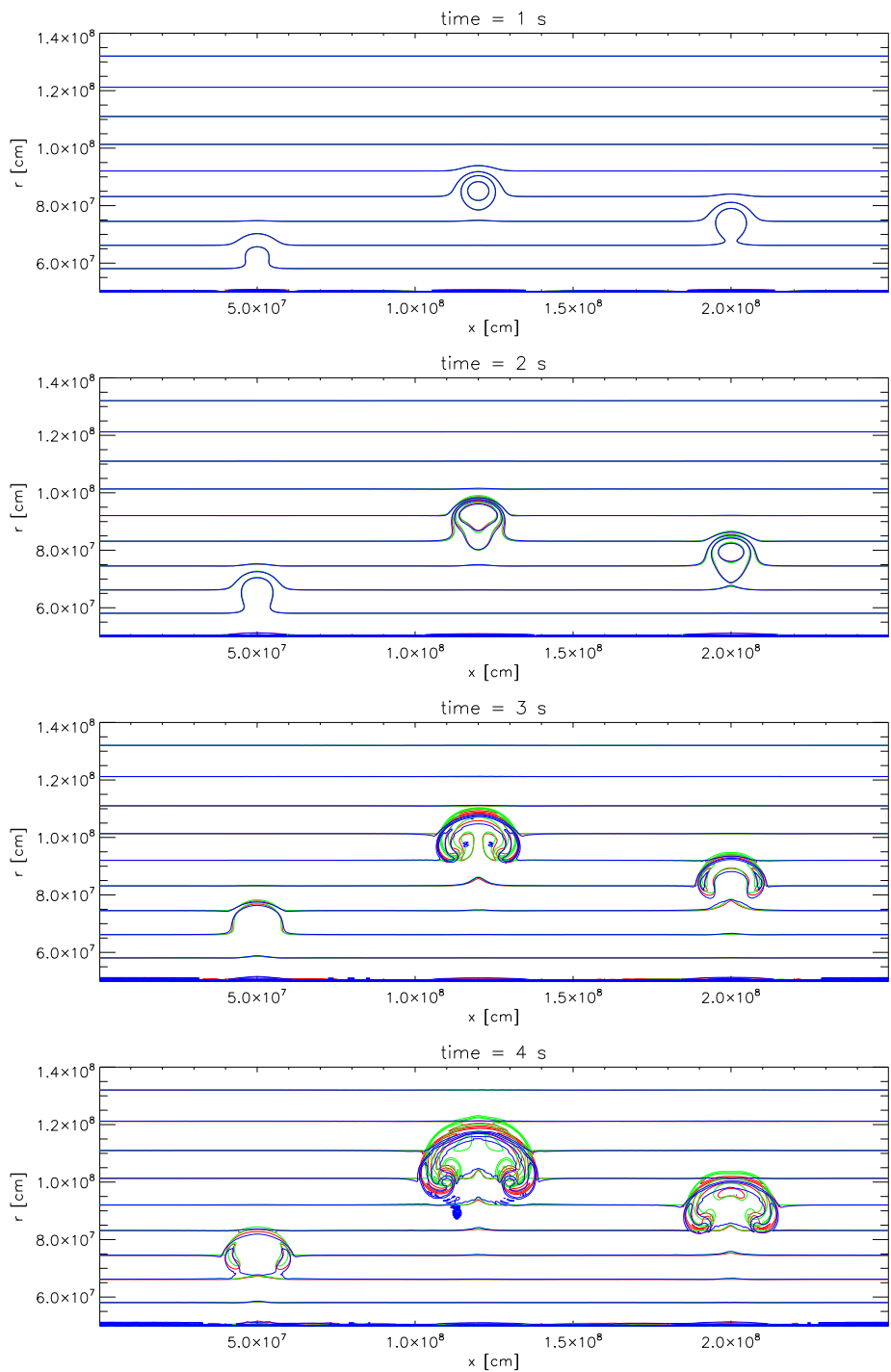}
\epsscale{1.0}
\end{center}
\caption{Second test case: temperature contours for the low Mach
number (green), unsplit (red), and PPM (blue) solvers, shown at 1,
2, 3, and 4~s. Here, a heating source term gradually adds energy
at three points in the domain (see Eq.~\ref{eq:nonuniformheat_1})
during the first 2~s of evolution.  This gives rise to the three
buoyant plumes seen in the panels.  Contours span $10^8$~K to $8\times
10^8$~K, spaced every $5\times 10^7$~K. }
\label{fig:test_case_2}
\end{figure*}

In the second test, we specify three local regions of heating, designed
to mimic ``hot spots,'' but no heated layer as in the first
case. In this scenario, while both the horizontal average and the
local deviations from the horizontal average are non-zero,
the deviations are much larger than the average,
so the dominant effect is the local rather than horizontally averaged
atmospheric response to the heating.  
For the first two seconds the heating profile has the form 
\begin{equation}
\label{eq:nonuniformheat_1}
            H = H_0 \left \{ \sum_{i=1}^3
                    \frac{a_i}{2} \left( 1 + \tanh (2 - d_i/ \sigma_i) \right) \right \}
\end{equation}
where $d_i = \sqrt{(x-x_i)^2 + (r-r_i)^2}$ and
the amplitudes, $a_i$, locations, $(x_i,r_i)$, and widths, $\sigma_i$, of
the perturbations are given in Table~\ref{table:pert}.  
After the first two seconds $H$ is set to zero, and we continue to follow 
the evolution until $t=4$~s.  
Figure~\ref{fig:test_case_2} shows a time sequence from $t=1$ to $t=4$~s
of temperature contours in a subset of the domain spanning from $5 \times 10^7$cm
to $1.4 \times 10^8$cm high. 
The temperature is an independent variable in the compressible calculations,
but in the low Mach number model, for these examples, we evolve density and 
compute temperature from the equation of state using density and $p_0$.

At early times, the three methods agree very well; at later times they
diverge slightly.  The vertical speed of the bubbles appears greatest
with the low Mach number methodology followed by the unsplit
compressible formulation; the PPM generates the slowest bubbles.  We
note that the difference in height between the FLASH PPM and unsplit methods
is comparable to the difference between the unsplit and the low Mach
number bubble heights.  The precise reason for these differences is
not yet completely understood; however, they serve to underscore the
sensitivity of these flows and the difficulties in simulating them
accurately with either the compressible or low Mach number approach.

\begin{figure*}

\begin{center}
\epsscale{1.0}
\plotone{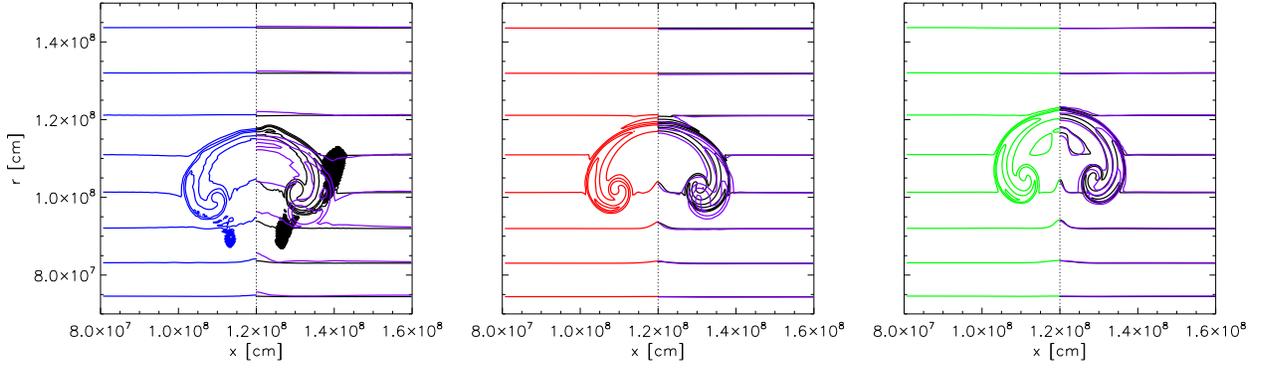}
\epsscale{1.0}
\end{center}
\caption{Resolution study for the second test case.  Each pane shows
the middle bubble from Figure~\ref{fig:test_case_2} only;  PPM results
are in the left pane, unsplit results are in the middle pane, and low Mach
number results are in the right pane.  Within each pane,  the left half is
identical to the left half of the bubble in Figure~\ref{fig:test_case_2},
with the same color scheme.   On the right half of each pane is a reflection
around the center line of the bubble of the comparable results but at both
a lower ($320 \times 448$) and higher ($1280 \times 1792$) resolution.
The low resolution is shown in purple and the high resolution is shown in black.}
\label{fig:test_case_2_res}
\end{figure*}

Figure~\ref{fig:test_case_2_res} shows a resolution study of this
second test case for each of the three methods.  The general trend one
observes is that the location of the bubble rises with increased
resolution.  However, we also notice that the low Mach number model
appears to converge to a solution at a lower resolution than either of
the compressible models.  This is likely due to the fact that
hydrostatic equilibrium is guaranteed in the low Mach number method by
the base state, while the compressible methods need considerable
resolution just to keep the background medium quiescent.  We also
notice the difficulty that the FLASH PPM method has at the higher
resolution, evidenced by the strong oscillations in temperature.  This
was also observed and discussed in paper I.

\begin{figure*}

\begin{center}
\epsscale{0.6}
\plotone{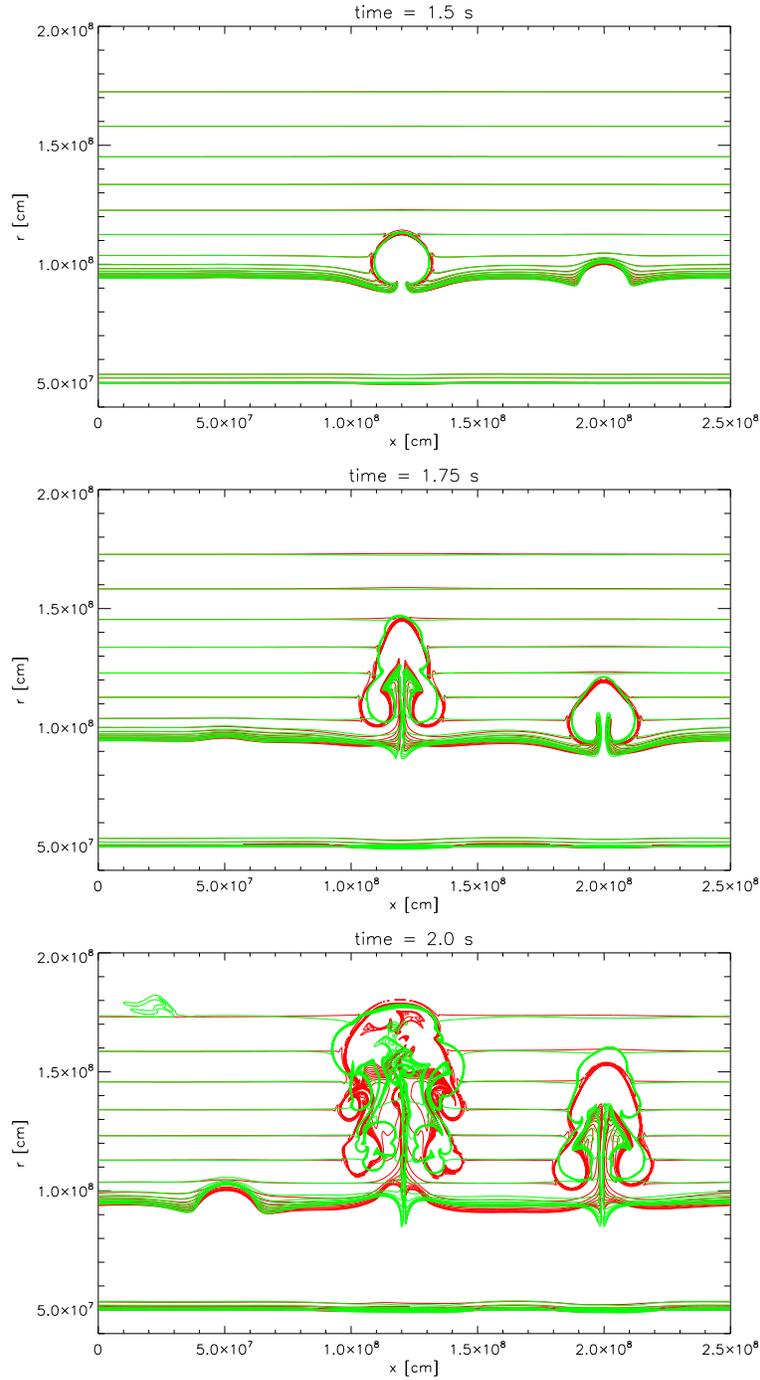}
\epsscale{1.0}
\end{center}

\caption{ Third test case: temperature contours for the low Mach
number (green) and unsplit (red) solvers.  Here, in addition to the
localized heating from the previous test (see Figure~\ref{fig:test_case_2}), 
there is a uniform heating layer centered at a
height of $7.5\times 10^7$~cm, as specified by
Eq.~\ref{eq:nonuniformheat_2}.  Contours span $10^8$~K to $8\times
10^8$~K, spaced every $5\times 10^7$~K. }
\label{fig:nonuniform_compare}
\end{figure*}

In the third test, we add the heated layer of the first test case
to the three ``hot spots'' of the second case, resulting in a case
for which the horizontal average of the heating is larger than the
perturbation from the average.  
The heating profile has the form 
\begin{equation}
\label{eq:nonuniformheat_2}
           H = H_0 \left \{ \exp(-(r-r_0)^2/W^2) + \sum_{i=1}^3
                    \frac{a_i}{2} (1 + \tanh (2 - d_i/ \sigma_i)) \right \}
\end{equation}
where $r_0 = 7.5\times 10^7$~cm and the amplitudes, $a_i$, locations,
$(x_i,r_i)$, and widths, $\sigma_i$, of the perturbations are as in
the previous case and are given in Table~\ref{table:pert}.  We apply
the heating source for 2~s.  As in the one-dimensional uniform heating
case, we place the heating layer a bit above the lower boundary, so as
to minimize contamination of the solution from lower boundary effects.
Figure~\ref{fig:nonuniform_compare} shows the temperature contours for
the unsplit compressible and low Mach number solver at $t=1.5$~s,
$t=1.75$~s and $t=2$~s.  We again notice excellent agreement between
the low Mach number and compressible results.  We do not expect the
exact shape of the rising bubbles to match precisely given the
extremely unstable nature of the bubble's surface, but there is
overall agreement between methods.  Once again the low Mach number
bubble rises slightly faster.  We do not show the FLASH PPM results here, as
the noise resulting from the dimensional splitting dominates the
solution.
We also note that at the final time, the low Mach number result shows a disturbance
in the upper left corner of the domain. This disturbance is a result of a spurious
wave generated at the outer boundary, which is not present in the compressible results 
because of the different treatment of the outer boundary condition.  The
question of how best to represent the outer boundary of the white dwarf in
a low Mach number calculation is still an open research question; at this point we note that while the 
high-velocity disturbance does restrict the time step, it does not appear to impact
the solution in the primary region of interest.  We also expect that
more realistic heating profiles in three-dimensional geometries
will not generate the disturbances in the outer boundary that we have observed 
in this final test case. 

\begin{figure*}
\begin{center}
\epsscale{0.6}
\plotone{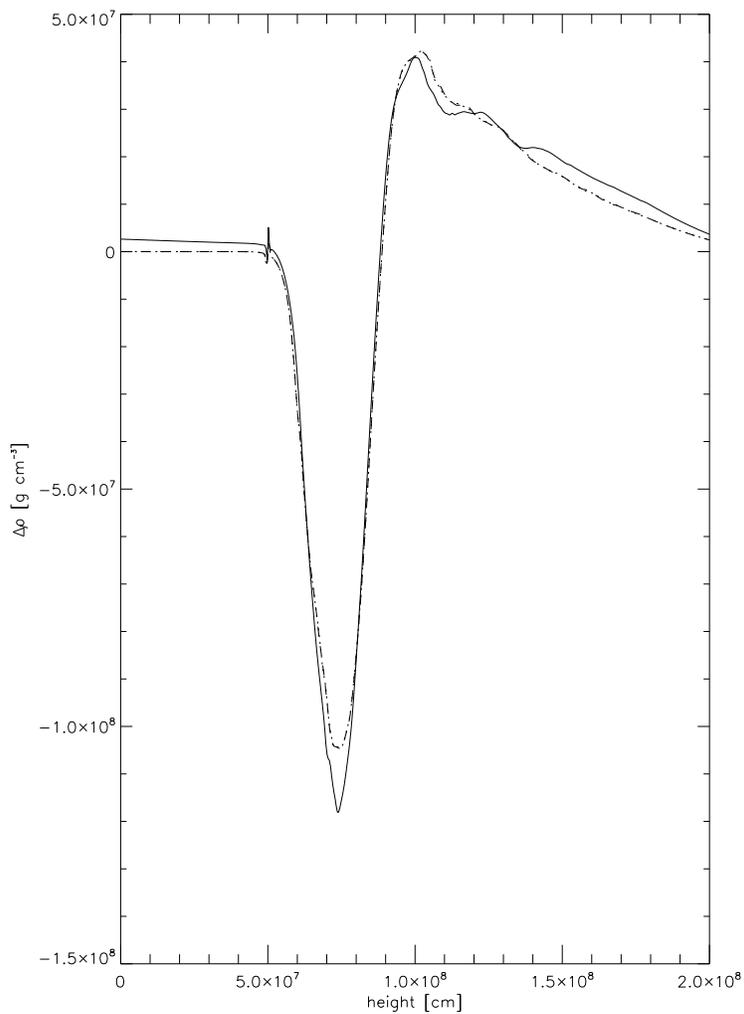}
\epsscale{1.0}
\end{center}
\caption{ Third test case: horizontal average of the difference between density at $t=2$ and
density at $t=0$.   The dotted line shows the low Mach number results; the solid line shows
results using the unsplit compressible formulation. The discrepancy between
the two solutions on the left is due to the difficulties in prescribing 
an accurate hydrostatic lower boundary for the fully compressible calculation.
Overall, however, the two algorithms agree well in the average response of
the atmosphere to the heating.
}
\label{fig:compare-test3-avg}
\end{figure*}

Figure~\ref{fig:compare-test3-avg} shows the horizontal average of the
difference between density at $t=2$s and $t=0$s for the low Mach
number and unsplit compressible results shown in
Figure~\ref{fig:nonuniform_compare}.  On the left of this plot we see
that the compressible solution has an increase in density over time
below the level of the applied heating;  this was also present in the
first test case, although to a lesser degree.  This error results
from the difficulty in prescribing an accurate hydrostatic lower boundary.
We note, however, that this is less than a 0.1\% relative
error in the density at this boundary.  At approximately the center
of the heated layer ($7.5 \times 10^7$~cm) each calculation shows a
negative average density variation.
The small difference here between the compressible and low Mach number
results here may also be a product of the lower boundary condition.
Overall, however, the two algorithms agree well in the average response of 
the atmosphere to the heating.

Finally, since computational efficiency as well as accuracy is
necessary for successful long-time integration, we comment on the
relative efficiency of the low Mach number algorithm.  For the second
test case, for example, to evolve the state to 4.0~s, the unsplit
compressible method took 14272 time steps while the low Mach number
algorithm took only 233 time steps.  
Evolving to just 2.0~s on a single processor
(2.8 GHz Intel Xeon) using the Intel 9.0 compilers (with {\tt -O3
-ipo} optimization flags) took 71.76 hours with the FLASH PPM solver.
FLASH was setup to use 32$\times$ 32 zone blocks for greatest
efficiency.  By comparison, the low Mach number method took only 0.46
hours---a factor of over 150 speed-up.  
The unsplit method takes the same timestep as the
PPM algorithm but is approximately a factor of two slower, due
to the additional work per timestep.

For the final test case, evolving the solution to 1.5~s took
6532 timesteps for the unsplit method and 749 time steps for the low Mach
number algorithm.  Already by this point in the calculation, the spurious
disturbance at the outer boundary of the star in the low Mach number 
algorithm is impacting the time step, decreasing the relative advantage of 
the low Mach number algorithm.   As noted above, we expect this not
to be the case for more realistic ignition scenarios.  However, this case 
points out that to achieve the gains in efficiency possible with a low Mach 
number model one must successfully address this issue.

\section{Conclusions}
\label{sec:conclusions}

We have introduced a new algorithm for evolving low Mach number flows
in the presence of local and large-scale heating.  By contrast with
the previous low Mach number model, this new model allows time
variation of the base state in order to account for atmospheric
expansion due to large-scale heat sources.  The time evolution of the
base state must be calculated at each time step in addition to the
local dynamics.  Numerical comparisons of low Mach number simulations
with simulations using a fully compressible code demonstrate that the
low Mach number algorithm with a time-dependent base state can
accurately capture the hydrostatic adjustment of an atmosphere as well
as local dynamics in response to large- and small-scale heat release.

Our long-term goal is to develop the capability for full star
simulation using the new low Mach number approach.  The fundamental
low Mach number approach has been validated with a number of
simplified test cases, but further development is necessary to begin
to perform detailed physics investigations of ignition and other problems
of interest.  This
development will include extension to three dimensions with adaptive
griding, radial representation of gravity and the base state within
the three-dimensional setting, non-constant gravity, and the
calculation of internal heating due to reaction networks.  

The tests presented above are quite demanding, and provided a
challenge to both the low Mach number and the compressible solvers.
The energy generation and resulting temperature/density contrasts
during the convective phase of an SN Ia are much smaller.  Based on
the agreement demonstrated on these difficult tests, we are confident
that the low Mach number hydrodynamics method will be a useful and
efficient tool in exploring the problem of SNe~Ia ignition.
In addition, this algorithm is applicable to a wide range of problems outside 
of our target application (SNe Ia), including Type I X-ray bursts, classical
novae, and convection in stars.

\acknowledgments

We especially thank Stan Woosley for numerous discussions and
interactions.  We thank Jonathan Dursi for helpful comments on this
paper.  The compressible calculations presented here used portions of
the FLASH Code (version 2.5), developed in part by the DOE-supported
ASC/Alliance Center for Astrophysical Thermonuclear Flashes at the
University of Chicago.  Some calculations made use of resources of the
National Energy Research Scientific Computing Center, which is
supported by the Office of Science of the U.S. Department of Energy
under Contract No.\ DE-AC02-05CH11231.  This work was supported by the
Applied Mathematics Program of the DOE Office of Mathematics,
Information, and Computational Sciences under the U.S. Department of
Energy under contract No.\ DE-AC02-05CH11231, by DOE grant No.\
DE-FC02-01ER41176 to the Supernova Science Center/UCSC, and by the
NASA Theory Program (NAGW-5-12036/NNG05GG08G).

\appendix
\section{Construction of Advective Updates}
\label{sec:appendix}

Consider the construction of an advective update in the form
$\left[ \nabla \cdot (s \V) \right]^{\nph}$, given the cell-centered
velocity field $\ubold^n = (u,v)$, an edge-based
velocity field, $\V = (\Vx,\Vr) $ and a cell-centered scalar, $s$.  For 
simplicity we will present the construction in two dimensions, 
although extension to three dimensions is straightforward and is
given in detail in \citep{almgren-iamr}.

We first extrapolate $s$ from cell centers at $t^n$ to edges
at $ t^{\nph}$ using a second-order
Taylor series expansion in space and time.  The time derivative
is replaced using the evolution equation for $s$.
If, for example, $s_t = -\nabla \cdot (s \V) = -\V \cdot \nabla s - s \nabla \cdot \V$, then
\begin{eqnarray*}
\tilde{s}_{i+\half,j}^{L} &\approx& s_{i,j} + \frac{\dx}{2} s_x + \frac{\dt}{2} s_t  \\
&=& s_{i,j} + (\frac{\dx}{2} - u_{i,j} \frac{\dt}{2})
(s^{\mathrm{lim}}_x)_{i,j} + \frac{\dt}{2} (-(\widehat{v s_r})_{i,j} 
- s_{i,j} (\mathrm{V}_x^x + \mathrm{V}_r^r)_{i,j} 
\end{eqnarray*}
extrapolated from $(i,j)$, and
\begin{eqnarray*}
\tilde{s}_{i+\half,j}^{R} &\approx& s_{i+1,j} - \frac{\dx}{2} s_x + \frac{\dt}{2} s_t  \\
&=& s_{i+1,j} + (-\frac{\dx}{2} - u_{i+1,j} \frac{\dt}{2})
(s^{\mathrm{lim}}_x)_{i+1,j} + \frac{\dt}{2} (-(\widehat{v s_r})_{i+1,j} 
- s_{i+1,j} (\mathrm{V}_x^x + \mathrm{V}_r^r)_{i+1,j} 
\end{eqnarray*}
extrapolated from $(i+1,j)$.
In evaluating these terms the first derivatives normal to the face (in this
case $ s_x^{\mathrm{lim}}$) are evaluated using a monotonicity-limited fourth-order
slope approximation \citep{colella1985}.  
The construction of the transverse derivative terms ($\widehat{v s_y}$ in this case)
are given in detail in \citep{almgren-iamr}.
Analogous formulae are used to predict 
values for $\tilde{s}_{i,j+\half}^{T/B}$ and 
$\tilde{s}_{i,j-\half}^{T/B}$ at the other cell edges.  

Upwinding is used to determine $\tilde{s}$ at each edge as follows:
\[ \tilde{s}_{\iphj} = \left\{\begin{array}{lll}
 \tilde{s}_{\iphj}^L 
& \mbox{if $\mathrm{V}_{\iphj}^{x} > 0$} \\
\half (\tilde{s}_{\iphj}^L + \tilde{s}_{\iphj}^R)  
& \mbox{if $\mathrm{V}_{\iphj}^{x} = 0$} \\
 \tilde{s}_{\iphj}^R
& \mbox{if $\mathrm{V}_{\iphj}^{x} < 0$} \end{array} \right.
\]
and similarly for defining $\tilde{s}_{\ijph}$ using $\mathrm{V}^r$.
Finally, we define the conservative update term,
\[ 
[\nabla \cdot (s \V) ]_{i,j}^{\nph} = 
  (\mathrm{V}_{\iphj}^x \tilde{s}_\iphj - \mathrm{V}_{\imhj}^x \tilde{s}_{\imhj})
+ (\mathrm{V}_{\ijph}^r \tilde{s}_\ijph - \mathrm{V}_{\ijmh}^r
\tilde{s}_{\ijmh}) \enskip .
\]

The construction of ${p_0}_{j+\myhalf}^{\nph}$ and ${\rho_0}_{j+\myhalf}^{\nph}$
in \textbf{Step 3} is similar but not identical to the above procedure.  Here we
make no reference to $\ubold^n$.  Rather
\[
{p_0}_{j+\half}^{B} = {p_0}_{j}^n + (\frac{\dr}{2} - {w_0}_j \frac{\dt}{2}) ({p_0}^{\mathrm{lim}}_r)_{j} 
\]
extrapolated from $j$, and
\[
{p_0}_{j+\half}^{T} = {p_0}_{j+1}^n - (\frac{\dr}{2} + {w_0}_{j+1} \frac{\dt}{2}) ({p_0}^{\mathrm{lim}}_r)_{j+1} 
\]
extrapolated from $j+1$.  Here ${w_0}_j = 1 / 2 ( {w_0}_{j+\half} + {w_0}_{j-\half} )$.
Upwinding then determines ${p_0}_{j+\myhalf}^{\nph}$:
\[ {p_0}_{j+\half}^{\nph} = \left\{\begin{array}{lll}
 {p_0}_{j+\half}^B 
& \mbox{if ${w_0}_{j+\half} > 0$} \\
\half ({p_0}_{j+\half}^T + {p_0}_{j+\half}^B)  
& \mbox{if ${w_0}_{j+\half} = 0$} \\
 {p_0}_{j+\half}^T
& \mbox{if ${w_0}_{j+\half} < 0$} \end{array} \right.
\] 

The evolution equation for $\rho_0$ differs from that for $p_0$ so the
construction of ${\rho_0}_{j+\myhalf}^{\nph}$ differs slightly from 
that for ${\rhozero}_{j+\myhalf}^{\nph}$.
\[
{\rhozero}_{j+\half}^{B} = {\rhozero}_j^n + 
     (\frac{\dr}{2} - {w_0}_{j} \frac{\dt}{2}) ({\rhozero}^{\mathrm{lim}}_r)_{j}
    - \frac{\dt}{2} {\rhozero}_j^n ({w_0}_{j+\half} - {w_0}_{j-\half})
\]
extrapolated from $j$, and
\[
{\rhozero}_{j+\half}^{T} = {\rhozero}_{j+1}^n - (\frac{\dr}{2} + {w_0}_{j+1} \frac{\dt}{2}) ({\rhozero}^{\mathrm{lim}}_r)_{j+1}
    - \frac{\dt}{2} {\rhozero}_{j+1}^n ({w_0}_{j+3/2} - {w_0}_{j+\half})
\]
extrapolated from $j+1$.
The upwinding procedure is the same.

\clearpage


\clearpage

\begin{table*}
\begin{center}
\caption{\label{table:pert} Location of heating sources for non-uniform heating terms, 
Eq.~\ref{eq:nonuniformheat_1} and Eq.~\ref{eq:nonuniformheat_2}.}
\begin{tabular}{rrrrr}
\tableline
\tableline
\multicolumn{1}{c}{$i$} & \multicolumn{1}{c}{$a_i$} & \multicolumn{1}{c}{$x_i$} & \multicolumn{1}{c}{$r_i$} & \multicolumn{1}{c}{$\sigma_i$} \\
 & & \multicolumn{1}{c}{(cm)} & \multicolumn{1}{c}{(cm)} & \multicolumn{1}{c}{(cm)} \\
\tableline
1 & $0.00625$ & $5.0\times 10^7$ & $6.5\times 10^7$ & $2.5\times 10^6$ \\
2 & $0.01875$ & $1.2\times 10^8$ & $8.5\times 10^7$ & $2.5\times 10^6$ \\
3 & $0.01250$ & $2.0\times 10^8$ & $7.5\times 10^7$ & $2.5\times 10^6$ \\
\tableline
\end{tabular}
\end{center}
\end{table*}

\clearpage

\clearpage

\clearpage

\clearpage

\clearpage

\clearpage

\clearpage

\end{document}